\documentclass[cmex10,reqno]{amsart}

\usepackage{moreverb}
\usepackage{graphicx}

\newtheorem{thm}{Theorem}
\newtheorem{prob}{Problem}
\newtheorem{prop}{Proposition}
\newtheorem{rem}{Remark}
\newtheorem{assum}{Assumption}
\usepackage{amsmath}
\usepackage{amsfonts}
\newcommand{\re}{{\mathbb{R}}}
\newcommand{\dd}{{\mathrm{d}}}
\newcommand{\D}{{\mathcal{D}}}

\newcommand{\F}{{\mathcal{F}}}
\newcommand{\tF}{\widetilde{\mathcal{F}}}
\newcommand{\one}{{\mathbf{1}}}
\newcommand{\inter}{{\mathrm{int}}}
\newcommand{\ipr}[2]{\langle #1, #2 \rangle}

\newcommand{\rev}[1]{{#1}}
\newcommand{\ind}[1]{}
\DeclareMathOperator*{\argmin}{arg\,min}

\begin{document}

\title[Monotone Smoothing Splines Using General Linear Systems]{Monotone Smoothing Splines Using\\ General Linear Systems%
\footnote{Asian Journal of Control, Vol.~15, No.~2, pp.~461--468, March 2013.}%
\footnote{Published online 25 June 2012 in Wiley Online Library (wileyonlinelibrary.com) DOI: 10.1002/asjc.557}%
}
\author[M. Nagahara]{Masaaki Nagahara}
\author[C. F. Martin]{Clyde F.~Martin}
\address{M.~Nagahara is with 
Graduate School of Informatics, Kyoto University,
Kyoto 606-8501, JAPAN. (nagahara@ieee.org).
C.~F.~Martin is with
Department of Mathematics \& Statistics, Texas Tech University,
Texas, USA.(clyde.f.martin@ttu.edu).
M.~Nagahara is the corresponding author.
}


\maketitle

\begin{abstract}
In this paper, a method is proposed to solve the problem of
monotone
\rev{\ind{1-1}\ind{3-1}
smoothing
}
splines using general linear systems.
This problem, also called monotone control theoretic splines, 
has been solved only when
the curve generator is modeled by the second-order integrator,
but not for other cases. 
The difficulty in the problem is that the monotonicity constraint should be
satisfied over an interval which has the cardinality of the continuum.
To solve this problem, we first formulate the problem as a semi-infinite
quadratic programming, and then we adopt a discretization technique
to obtain a finite-dimensional quadratic programming problem.
It is shown that the solution of the
finite-dimensional problem always satisfies the infinite-dimensional
 monotonicity constraint. It
is  also proved that the approximated solution converges to the
exact solution as the discretization grid-size tends to zero.
An example is presented to show the effectiveness of the proposed method.
\end{abstract}

\section{Introduction}
\label{sec:introduction} 
{\em Control theoretic splines} are either
interpolating or smoothing splines, depending on the constraints
and cost function, with a constraint written as a system of linear
differential equations~\cite{SunEgeMar00}. The spline curve
 is obtained as the output of a given linear system. 
This concept is generalization of the smoothing spline
by Wahba~\cite{KimWah71,Wah}
in that a richer class of smoothing curves can be obtained
relative to polynomial curves.
They have been proved to be useful in trajectory planning~\cite{EgeMar01,Rub+10}, 
mobile robots~\cite{TakMar04}, 
contour modeling of images~\cite{KanEgeFujTakMar08},
probability distribution estimation~\cite{Cha09,Cha10}, 
and so on. For more
applications and a rather complete theory of control theoretic
splines, see the book~\cite{EgeMar}.

On the other hand, {\em monotone} control theoretic splines are
also important in deriving a model or an estimation of a parameter
such as the growing rate of an individual~\cite{EgeMar00,EgeMar03,EgeMar}.
Also, there have been a few applications of monotone smoothing in the
statistical literature. The work of Ramsay~\cite{Ramsay} and the
discussion associated with the paper seems to be the first use. A
more recent paper by Kelly~\cite{Kelly} is  a nice application.

Monotonicity is achieved by adding a non-negative 
(or non-positive) derivative
constraint on the output of the target linear system over a time
interval. Since the interval is an infinite subset of
$[0,\infty)$, this constraint becomes infinite-dimensional, which
makes the problem difficult to solve. In~\cite{EgeMar00,EgeMar03},
this has been solved when the target linear system is a
second-order integrator $1/s^2$. However, for general cases, no
solution has been obtained.
More recently, the smoothing problem has been solved
with inequality constraints~\cite{BelBurPil09},
in which inequality constraints are defined only on sampling points,
and the constraints may be violated between sampling points.

In this paper, we propose a new method for constructing monotone
splines guaranteeing the inequality constraint on a whole time interval,
say $[0,T]$, with more general linear systems.
To solve this, we first
formulate the problem as a semi-infinite quadratic programming
problem~\cite{HetKor93,LopSti07}. Then we adopt a discretization
technique~\cite{HetGra90,TeoYan00} to reduce the
infinite-dimensional constraints to a finite dimensional
constraints. This discretization is executed by partitioning the
time interval into finitely many subintervals and estimating the
constraint on a finite grid. 
The finite-dimensional quadratic
programming problem can be solved with efficient algorithms~\cite{BoyVan}. 

We also discuss the limiting properties of the approximated
solution. Whereas the approximated solution is constrained on a
finite grid, we show that it satisfies the original
infinite-dimensional constraint provided   the grid size is
sufficiently small. We also prove that the approximated solution
converges to the exact solution as the discretization grid-size
tends to zero. By these properties, the proposed method can be
safely adopted for monotone control theoretic splines. We also
construct an illustrative   example to show the effectiveness of
our method.

The method proposed in this paper can be directly applied to
model predictive control (MPC),
in particular nonuniformly-sampled-data MPC with continuous-time
inequality constraints.
For the study, there are many researches;
MPC with uniform sampling and piecewise 
constant control~\cite{MagSca04,GruNesPan07},
with constraints only on the sampling points~\cite{Die-etal02},
and with nonuniformly sampling but unconstrained~\cite{SheCheSha02}.
To our knowledge, there is no method for nonuniformly-sampled-data MPC
with relatively general control inputs which guarantees continuous-time
constraints as we will consider in this paper.

The organization of this paper is as follows:
Section~\ref{sec:definition} defines
the problem of  monotone control theoretic splines.
In Section~\ref{sec:formulation}, we formulate the problem
as a semi-infinite quadratic programming.
Section~\ref{sec:discretization} is the main section,
where discretization approach is introduced and
the limiting properties are discussed.
Section~\ref{sec:computing} suggests a formula
for computing inner products used in our optimization.
A numerical example is illustrated in Section~\ref{sec:examples},
and a conclusion is made in Section~\ref{sec:conclusion}.

\subsection*{Notation}
In this paper, we use the following notation. 
$\re$, $\re^n$, and
$\re^{n\times n}$ are respectively the set of real numbers,
$n$-dimensional real vectors and $n\times n$ matrices. We denote
$L^2[0,T]$ by the Lebesgue space consisting of all square
integrable real functions on $[0,T]\subset\re$, endowed with the
inner product
\[
 \ipr{x}{y} := \int_0^T x(t)y(t) \dd t, \quad x, y \in L^2[0,T].
\]
For a matrix (a vector) $M$, $M^\top$ is the transpose of $M$.
For a vector $v=[v_1,\dots,v_n]^\top\in\re^n$, $v\preceq 0$ means $v_i\leq 0$ for $i=1,2,\dots, n$,
and for two vectors $v$ and $w$, $v\preceq w$ if $v-w\preceq 0$.
For a vector $v\in\re^n$,
\[
 \begin{split}
 \|v\|_1&:=\sum_{i=1}^n |v_i|,\\
 \|v\|_\infty &:= \max_{i=1,\dots,n} |v_i|.
 \end{split}
\]
\section{Monotone Smoothing Splines}
\label{sec:definition}
In this paper, we consider the following problem of monotone smoothing splines:
\begin{prob}[Monotone smoothing splines]
\label{prob:monotone}
Given a linear time-invariant system
\[
 \dot{x} = Ax + Bu, \quad y = Cx,~
 \rev{\ind{3-3}
  x(0) = x_0 \in\re^n},
\]
where
\rev{\ind{3-2} 
$x(t)\in\re^n$ is the state vector,
$u(t)\in\re$ is the control input,
$y(t)\in\re$ is the plant output},
$A\in\re^{n\times n}$, $B\in\re^{n\times 1}$, $C\in\re^{1\times n}$,
and also given a data set
\[
\D:=\bigl\{(t_1, \alpha_1), (t_2, \alpha_2),\dots, (t_m, \alpha_m)\bigr\},
\]
with time instants $0<t_1<t_2<\dots<t_m=:T$ and 
\rev{\ind{1-10}
noisy
}
data $\alpha_1,\dots, \alpha_m\in \re$
on the time instants,
find the control $u\in L^2[0,T]$ and the initial condition $x(0)=x_0\in\re^n$
that minimizes the following cost function:
\begin{equation}
 J(u) := \lambda \int_0^T u(t)^2 \dd t 
  + \sum_{i=1}^m w_i \bigl(y(t_i) - \alpha_i\bigr)^2,
 \label{eq:cost}
\end{equation}
where $\lambda$ and $w_i$'s are given positive numbers (weights),
subject to the monotonicity constraint
\begin{equation}
 \dot{y}(t) \geq 0,\quad \forall t \in [0,T].
 \label{ieq:monotone}
\end{equation}
\end{prob}

The cost function \eqref{eq:cost} considers elimination of Gaussian noise on the data $\{\alpha_1,\dots,\alpha_m\}$
and limitation on the $L^2$ norm of the control $u$.
This formulation is an extension of Wahba's smoothing spline~\cite{KimWah71,Wah}.

This problem has been solved in~\cite{EgeMar00,EgeMar03} only when
$C(sI-A)^{-1}B=1/s^2$.
However,  for other cases, the problem has not been solved.

\begin{rem}
One might think that the problem can be solved via a technique of {\em model predictive control} (MPC).
The difficulty in using MPC is that we have to treat the constraint 
\eqref{ieq:monotone}
which is infinite dimensional.
If we discretize this constraint as
$\dot{y}(t_i)\geq 0$ $i=1,2,\dots,m$
to use a standard MPC formulation,
the original constraint \eqref{ieq:monotone} will be guaranteed only on the sampling points
$t_1,\dots,t_m$, not on the whole $[0,T]$.
As mentioned in Section 1, there is no method for the problem
given in \eqref{eq:cost} and \eqref{ieq:monotone} within the framework of MPC.
\end{rem}

\section{Formulation by Semi-Infinite Quadratic Programming}
\label{sec:formulation}
We here formulate the problem given in the previous section by semi-infinite quadratic programming.
We first assume that we search the optimal control $u$ in the following subspace of $L^2[0,T]$
as taken in~\cite{SunEgeMar00}:
\[
 V_m := \biggl\{u\in L^2[0,T]: u = \sum_{i=1}^m \eta_i \phi_{t_i},~ \eta_i \in \re\biggl\},
\]
where $\phi_t\in L^2[0,T]$ is defined by
\begin{equation}
 \phi_{t}(\tau) := \begin{cases} ~Ce^{A(t-\tau)}B,&\quad t>\tau,\\ ~0,&\quad t\leq \tau.\end{cases}
 \label{eq:phi}
\end{equation}
This characterization is based on the 
{\em representer theorem}~\cite{KimWah71,Wah,SchSmo}
for the regularized cost functional \eqref{eq:cost}
with the kernel $k(t,\tau)=\ipr{\phi_t}{\phi_\tau}$;
the optimal estimation $y(t)$ is represented by
\[
 y(t) = \sum_{i=1}^m \eta_i k(t,t_i) = \sum_{i=1}^m \eta_i\ipr{\phi_t}{\phi_{t_i}}.
\]
One can consider other base functions $\{\phi_i\}_{i=1}^M$ than \eqref{eq:phi}
and define the subspace $V_M:=\text{span}\{\phi_i\}_{i=1}^M$ where $M\gg m$.
This is related to {\em basis pursuit denoising}~\cite{CheDonSau98},
with which we are not concerned in this paper.

By assuming that the control $u$ is in the subspace $V_m$, the cost function 
\eqref{eq:cost} can be described as~\cite{SunEgeMar00}
\begin{equation}
 J\left(\sum_{i=1}^m \eta_i \phi_{t_i}\!\right) = \frac{1}{2}\theta^\top P\theta + q^\top \theta + \alpha^\top W \alpha,
  \label{eq:cost2}
\end{equation}
where $\theta:=[\eta_1,\dots,\eta_m,x_0^\top]^\top$, $\alpha:=[\alpha_1,\dots, \alpha_m]^\top$, and
\[
 \begin{split}
 P &:= \begin{bmatrix} 2(\lambda I + GW)G & 2GWF \\ 2F^\top WG & 2F^\top WF \end{bmatrix},\quad
 q := -2\bigl[G ~~ F\bigr]^\top W\alpha,\\
 G &:= \begin{bmatrix}
        \ipr{\phi_{t_1}}{\phi_{t_1}}&\ipr{\phi_{t_2}}{\phi_{t_1}}&\ldots&\ipr{\phi_{t_m}}{\phi_{t_1}}\\
        \vdots & \vdots & \ddots & \vdots\\
        \ipr{\phi_{t_1}}{\phi_{t_m}}&\ipr{\phi_{t_2}}{\phi_{t_m}}&\ldots&\ipr{\phi_{t_m}}{\phi_{t_m}}\\
    \end{bmatrix},\quad
 F := \begin{bmatrix}Ce^{At_1}\\\vdots\\Ce^{At_m}\end{bmatrix},\\
 W &:= \text{diag}\left\{w_1,\dots,w_m\right\}.
 \end{split}
\]

Let $\dot{\phi}_t(\tau)$ be the derivative of $\phi_t(\tau)$, that is,
\begin{equation}
 \dot{\phi}_t(\tau) := \dfrac{\partial{\phi_t(\tau)}}{\partial t} 
  = \begin{cases}CAe^{A(t-\tau)}B,\quad \text{if}~~ t>\tau,\\ 0,\quad \text{if}~~ t\leq\tau.\end{cases}
  \label{eq:dphi}
\end{equation}
By using this, the derivative $\dot{y}(t)$ can be calculated as 
$\dot{y}(t)=\Phi(t)^\top\theta$,
where
\[
 \Phi(t) := \begin{bmatrix}\ipr{\dot{\phi}_t}{\phi_{t_1}}\\\vdots\\\ipr{\dot{\phi}_t}{\phi_{t_m}}\\ e^{A^\top t}A^\top C^\top\end{bmatrix}
    +CB \begin{bmatrix}\phi_{t_1}(t)\\\vdots\\\phi_{t_m}(t)\\0\end{bmatrix}.
\]
The constraint \eqref{ieq:monotone} now becomes $\Phi(t)^\top\theta\geq 0$ for all $t\in[0,T]$.
In summary, Problem~\ref{prob:monotone} is formulated by the following:
\begin{prob}
\label{prob:semiinf}
Find $\theta\in\re^{m+n}$ that minimizes
\begin{equation}
 f(\theta) = \frac{1}{2}\theta^\top P \theta +q^\top\theta,\label{eq:quad}
\end{equation}
subject to
\begin{equation}
  g(\theta,t)=-\Phi(t)^\top\theta\leq 0, \quad \text{for all }t\in[0,T].
  \label{ieq:semiinf}
\end{equation}
\end{prob}
\rev{\ind{1-3}
Note that the term $\alpha^\top W \alpha$ in \eqref{eq:cost2}
is omitted in \eqref{eq:quad} since this term
is constant (independent of $\theta$)
and does not change the optimal solution.
}
Problem~\ref{prob:semiinf} is  a semi-infinite quadratic programming
problem.
By the optimal solution $\theta^\ast$ to this problem, 
we obtain the optimal control input $u^\ast$ by
\[
 u^\ast(t) = \sum_{i=1}^m \theta^\ast[i]\phi_{t_i}(t),
\]
where $\theta^\ast[i]$ is the $i$-th element of $\theta^\ast$.

\section{Discretization Approach to Monotone Splines}
\label{sec:discretization}
\subsection{Discretization}
The difficulty in Problem~\ref{prob:semiinf} is that the inequality constraint
\eqref{ieq:semiinf} is infinite dimensional.
We here introduce an approximation technique for such an infinite-dimensional constraint.
To reduce this to a finite dimensional one, 
we adopt the technique of discretization~\cite{HetGra90,TeoYan00}.

We first divide the interval $[0,T]$ into $M$ subintervals:
$
 [0,T] = [T_0,T_1]\cup \dots \cup (T_{M-1}, T_M],
$
where $\{T_i\}$ is the discretization grid satisfying
$
 0 =T_0 < T_1 < \dots < T_M = T.
$
Then we evaluate the function $g(\theta, t)$ in \eqref{ieq:semiinf} on the grid points
$t=T_0,T_1,\dots,T_M$.
To guarantee the constraint \eqref{ieq:semiinf}, we choose real numbers $\epsilon>0$ and $r>0$
and adopt the following constraints:
\begin{equation}
 g(\theta, T_0) \leq -\epsilon,\dots,g(\theta, T_M) \leq -\epsilon,~
 -r\one \preceq \theta \preceq r\one
 \label{ieq:finite}
\end{equation}
where $\one=[1,1,\dots,1]^\top\in\re^{m+n}$.
The second inequality is component wise (see Notation in Section 1), 
and means $\|\theta\|_\infty\leq r$.
The finite-dimensional constraints \eqref{ieq:finite}
can be represented as a matrix inequality,
\begin{equation}
 H\theta \preceq v,
 \label{eq:const}
\end{equation}
where
\[
\begin{split}
H &:= \bigl[\begin{array}{ccccc}\Phi(T_0)&\ldots&\Phi(T_M)&I&-I\end{array}\bigr]^\top,\\
v &:= \bigl[\begin{array}{ccccc}-\epsilon&\ldots&-\epsilon&r\one^\top&r\one^\top\end{array}\bigr]^\top.
\end{split}
\]

In summary, the problem of monotone control theoretic splines
for arbitrarily linear time-invariant system $\{A, B, C\}$
can be approximately described by the following quadratic programming:
find $\theta\in\re^{m+n}$ which minimizes \eqref{eq:quad}
subject to \eqref{eq:const}.
This is a standard quadratic programming
and can be efficiently solved by numerical softwares such as MATLAB\@.
\begin{rem}
We can also include equality constraints into our optimization.
For example, if we want to have $y(T)=1$ and $\dot{y}(T)=0$,
then our constraints are represented by
\[
 \begin{split}
 \bigl[\ipr{\phi_T}{\phi_{t_1}},\dots,\ipr{\phi_T}{\phi_{t_m}}, Ce^{AT}\bigr]\theta &= 1,\\
 \bigl[\ipr{\dot{\phi}_T}{\phi_{t_1}},\dots,\ipr{\dot{\phi}_T}{\phi_{t_m}}, CAe^{AT}\bigr]\theta &= 0.
\end{split}
\]
These constraints are finite dimensional and can be included easily in our quadratic programming.
In general, equality constraints on the output $y$ and the derivative $\dot{y}$ on time 
\rev{\ind{1-5}
instants in $[0,T]$
can be represented by linear
constraints as above.
}
\end{rem}
\begin{rem}
\rev{\ind{1-14}
Splines with another constraint than monotonicity such as concavity, i.e., $\ddot{y}(t)\leq 0$,
can be also solved by the same method.
}
\end{rem}

\subsection{Convergence analysis}
Put $N:=m+n$.
Define feasible sets $\F$ for the original optimization
and $\tF(M,\epsilon, r)$ for the approximation
respectively by
\begin{equation}
  \F :=  \bigl\{ \theta\in\re^N: g(\theta,t)\leq 0,~ t\in[0,T]\bigr\},
  \label{eq:F}
\end{equation}
\begin{equation}
  \tF(M, \epsilon, r) :=
    \big\{
        \theta \in \re^N: g(\theta, t) \leq -\epsilon,~
   t\in \{T_0,\dots,T_M\},~ \|\theta\|_\infty\leq r
    \big\}.
  \label{eq:tF}
\end{equation}
To prove our first proposition, we assume the following:
\begin{assum}
\label{ass:lipschitz}
The function $\Phi(t)$ is Lipschitz continuous, that is,
there exists a real number $\mu>0$ such that
$\|\Phi(t)-\Phi(s)\|_1 \leq \mu |t-s|$ for all $t,s\in[0,T]$.
\end{assum}
\begin{rem}
Systems whose relative degree is higher than or equal to 2 satisfy the above assumption.
First, by the definition of $\dot{\phi}_t(\tau)$
in \eqref{eq:dphi},
we can say that for 
\rev{\ind{1-6}
$j=1,2,\dots,m$},
$\frac{d}{dt}\ipr{\dot{\phi}_t}{\phi_{t_j}}$ is bounded on $[0,T]$,
and hence $\ipr{\dot{\phi}_t}{\phi_{t_j}}$ is Lipschitz on $[0,T]$.
Also, the function $e^{A^\top t}A^\top C^\top$ is Lipschitz.
These facts and $CB=0$ (since the relative degree $\geq 2$) result in that $\Phi(t)$ is Lipschitz.
\end{rem}
\rev{\ind{1-8}
Define the discretization grid-size by
\[
I_{\max}(M):=\max\bigl\{\left|T_i-T_{i-1}\right|:{i=1,2,\dots,M}\bigr\}.
\]
}
We then have the following proposition:
\begin{prop}
\label{prop1}
\rev{\ind{1-7}\ind{1-8}
Suppose that Assumption~\ref{ass:lipschitz} is satisfied.
Suppose also that
$M$, $\epsilon$ and $r$ are chosen such that
\begin{equation}
 I_{\max}(M) \leq \frac{\epsilon}{r\mu}.
 \label{ieq:M}
\end{equation}
Then we have $\tF(M, \epsilon,r)\subseteq\F$.
}
\end{prop}
{\it Proof.~~~}
Let $\theta\in\tF(M,\epsilon,r)$.
\rev{\ind{1-8}
By the definition of $\tF(M,\epsilon,r)$
in \eqref{eq:tF}},
$\theta$ satisfies
$g(\theta, T_i)\leq -\epsilon$, $i=0,1,\dots,M$.
Then, for any $t\in[0,T]$, there exists $i\in\{1,\dots,M\}$ such that $t\in[T_{i-1}, T_i]$.
By Assumption~\ref{ass:lipschitz} and the condition $\|\theta\|_\infty\leq r$ 
in 
\rev{\ind{1-8}
\eqref{eq:tF}}, 
we have
\[
 g(\theta, t) - g(\theta, T_i) = -\left(\Phi(t)-\Phi(T_i)\right)^\top\theta \leq r\mu (T_i-t).
\]
By this inequality, we have
\[
 g(\theta, t) \leq r\mu(T_i-t) + g(\theta, T_i)
 \leq r\mu I_{\max}(M) - \epsilon
 \leq 0.
\]
The last inequality is due to \eqref{ieq:M}.
Thus, $\theta \in \F$ and hence $\tF(M, \epsilon,r)\subseteq\F$.
\hfill$\Box$

By this proposition, we can guarantee the constraint \eqref{ieq:semiinf}
by searching the parameter $\theta$ in the finite-dimensional feasible set
$\tF(M, \epsilon,r)$ provided that the number $M$ is sufficiently large
to satisfy \eqref{ieq:M}.

Next, we analyze the limiting property of the approximated solution
when the discretization grid size tends to zero.
To prove the property, we assume the following:
\begin{assum}
\label{ass:interior}
The interior of $\F$
\[
 \inter~\F := \bigl\{\theta\in\re^N: g(\theta,t)< 0, \text{\rm{ for all }}t\in[0,T]\bigr\}
\]
is nonempty.
\end{assum}
\begin{rem}
\rev{\ind{1-9}
Assumption~\ref{ass:interior}
}
is slightly restrictive.
A sufficient condition for this is that
$A$ is a Metzler matrix%
\footnote{A matrix $A$ is called a {\em Metzler matrix} if all the off-diagonal components are nonnegative.}
and $CA\succeq 0$ ($CA\not\equiv 0$).
By the positive system theory~\cite{FarRin},
$A$ is Metzler if and only if $e^{At}\succeq 0$ for all $t\in\re$.
Let $u\equiv 0$.
Then, we have $\dot{y}(t)=CAe^{At}x_0\geq 0$ for all $t\in\re$.
Since $CAe^{At}x_0$ is continuous on $\re$,
we have $\dot{y}(t)=CAe^{At}x_0>0$ for all $t\in[0,T]$,
if we choose sufficient large $x_0\succ 0$.
Therefore, $\inter~\F$ is nonempty in this case.
\end{rem}
Then we have the following theorem:
\begin{thm}
\label{thm:convergence}
\rev{\ind{1-7}
Suppose that Assumptions~\ref{ass:lipschitz} and~\ref{ass:interior}
are satisfied}.
Let $M(\epsilon,r)$ be the minimum value among all $M$'s satisfying
\eqref{ieq:M}.
\rev{\ind{2-1}
For $f(\theta)$ given in \eqref{eq:quad}
}, define
\[
 \theta^\ast := \rev{\argmin_{\theta \in \F}}\,f(\theta),\quad 
 \theta^\ast_{\epsilon,r} := \rev{\argmin_{\theta \in \tF(M(\epsilon,r),\epsilon,r)}} f(\theta).
\]
Then we have
$f(\theta^\ast_{\epsilon,r}) \rightarrow f(\theta^\ast)$
as $\epsilon\rightarrow +0$ and $r\rightarrow\infty$.
\end{thm}
{\it Proof.~~~}
Since $\inter~\F$ is nonempty, there exist $\bar{\theta}\in\inter~\F$.
It follows that there exists $\delta>0$ such that
\begin{equation}
g(\bar{\theta},t) \leq -\delta,\quad \forall t\in[0,T].
\label{eq:delta}
\end{equation}
Then, since $\F$ is a convex set, we have
$\theta_\xi:=\xi\bar{\theta}+(1-\xi)\theta^\ast \in \inter~\F$
for all $\xi \in (0,1]$.
Fix arbitrarily $\xi\in(0,1]$.
Since $f$ is a convex function, we have
\begin{equation}
 f(\theta_\xi) \leq \xi(\bar{\theta}) + (1-\xi)f(\theta^\ast) = f(\theta^\ast)+\xi \Delta,
 \label{eq:difference}
\end{equation}
where $\Delta:=f(\bar{\theta})-f(\theta^\ast)$.
Note that $\Delta\geq 0$ by the definition of $\theta^\ast$.
Also, for all $t\in[0,T]$, we have
\[
 g(\theta_\xi,t) = -\Phi(t)^\top\theta_\xi
   = \xi g(\bar{\theta},t)+(1-\xi)g(\theta^\ast,t)
   \leq -\xi\delta,
\]
where the last inequality is due to 
inequality \eqref{eq:delta}
and $g(\theta^\ast,t)\leq 0$ for $\theta^\ast\in\F$
defined in \eqref{eq:F}.
Let
$
 \bar{r}:=\max_{\xi\in(0,1]} \|\theta_\xi\|_\infty
$
and fix arbitrarily $r\geq \bar{r}$.
Then we have
$
 \theta_\xi \in \tF(M(\xi\delta,r),\xi\delta,r).
$
This gives the following inequality:
\begin{equation}
 f(\theta^\ast_{\xi\delta,r}) \leq f(\theta_\xi) \leq f(\theta^\ast) + \xi\Delta.
 \label{eq:thm-ieq1}
\end{equation}
On the other hand, by Proposition~\ref{prop1},
\rev{\ind{2-2}
we have $\F\supseteq\tF(M(\xi\delta,r),\xi\delta,r)$, and hence
}
\begin{equation}
 f(\theta^\ast) \leq f(\theta^\ast_{\xi\delta,r}).
 \label{eq:thm-ieq2}
\end{equation}
Combining \eqref{eq:thm-ieq1} and \eqref{eq:thm-ieq2} gives
$f(\theta^\ast) \leq f(\theta^\ast_{\xi\delta,r}) \leq f(\theta^\ast) + \xi \Delta$.
We thus have
\[
  \lim_{\substack{\epsilon\rightarrow +0\\ r\rightarrow\infty}} f(\theta^\ast_{\epsilon,r}) =
  \lim_{\substack{\xi\rightarrow +0\\ r\rightarrow\infty}} f(\theta^\ast_{\xi\delta,r}) = f(\theta^\ast).
\]
\hfill$\Box$
\begin{rem}
By Theorem~\ref{thm:convergence}, the semi-infinite programming we consider here is 
{\em weakly discretizable}~\cite{LopSti07}.
Note that this does not necessarily imply 
that $\theta_{\epsilon,r}^\ast\rightarrow\theta^\ast$.
\end{rem}
By Theorem~\ref{thm:convergence}, the proposed approximation method given in the previous section
can be safely adopted for monotone control theoretic splines.
\section{Computing Inner Product}
\label{sec:computing}
To compute the Grammian $G$ and the matrix $\Phi(t)$,
we have to compute the following inner products for fixed $s,~t\in [0,T]$:
\[
 \begin{split}
  \ipr{\phi_s}{\phi_t} &= \int_0^T\phi_s(\tau)\phi_t(\tau)\dd\tau,\\
  \ipr{\dot{\phi}_s}{\phi_t} &= \int_0^T\dot{\phi}_s(\tau)\phi_t(\tau)\dd\tau.
 \end{split}
\]
These values can be easily computed by matrix exponentials~\cite{Loa78}:
\[
 \begin{split}
  \ipr{\phi_s}{\phi_t} 
  &= B^\top\!e^{A^\top\!s}\biggl(\int_0^{h} e^{-A^\top\tau}C^\top Ce^{-A\tau}\dd\tau\biggr) e^{At}B\\
  &=: v_s^\top \left(F_{22}^\top F_{12}\right)v_t,
 \end{split}
\]
where $h:=\min(s,t)$, $v_{\tau} := e^{A\tau}B$ ($\tau=t,s$), and the matrices $F_{22}$ and $F_{12}$ are defined by
\[
  \begin{bmatrix}F_{11} & F_{12} \\ 0 & F_{22}\end{bmatrix} := \exp\left(\begin{bmatrix}A^\top & C^\top C\\ 0 & -A\end{bmatrix}h\right).
\]
The inner product $\ipr{\dot{\phi}_s}{\phi_t}$ is also obtained by
\[
 \begin{split}
  \ipr{\dot{\phi}_s}{\phi_t} &= v_s^\top \left(H_{22}^\top H_{12}\right) v_t,\\
  \begin{bmatrix}H_{11} & H_{12} \\ 0 & H_{22}\end{bmatrix} &:= \exp\left(\begin{bmatrix}A^\top & A^\top C^\top C\\ 0 & -A\end{bmatrix}h\right).
 \end{split}
\]

\section{Example}
\label{sec:examples}
\begin{figure}[t]
  \centering{
    \includegraphics[width=0.8\linewidth]{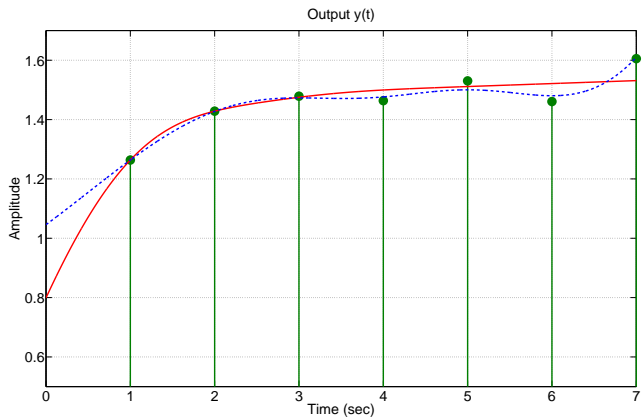}
  }
  \caption{Data (circles) and estimation $y(t)$ 
	by the proposed method (solid),
	and a conventional discretization method (dash)}
  \label{fig:reconstruction}
\end{figure}

\begin{figure}[t]
  \centering{
    \includegraphics[width=0.8\linewidth]{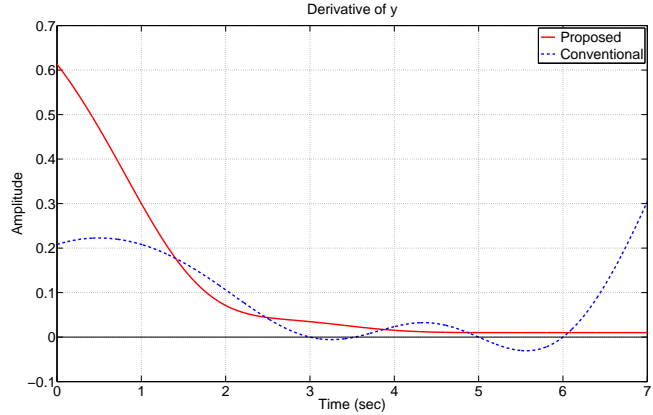}
  }
  \caption{Derivative of estimation $y(t)$  
		by the proposed method (solid),
		and a conventional discretization method (dash)}
  \label{fig:derivative}
\end{figure}

We here show an example of monotone control theoretic splines.
Assume that the system is given by
\begin{equation}
 P(s) = \frac{1}{s^2(s^2+1)}.
  \label{eq:Ps}
\end{equation}
Note that
\rev{\ind{1-13}
for the second order integrator $1/s^2$ the exact optimal solution
can be obtained~\cite{EgeMar00,EgeMar03}, but}
for higher order systems as in \eqref{eq:Ps},
there has been no method to construct monotone smoothing splines.
Let the original curve be $y_{\text{orig}}(t) = 1.5-e^{-t}$,
and the 
\rev{\ind{1-10}
noisy
}
data is obtained by
\[
 t_i=i,\quad \alpha_i = y_{\text{orig}}(t_i) + \varepsilon_i,\quad i=1,\dots,7,
\]
where $\varepsilon_i$ is an independently and identically distributed random variable
of the Gaussian distribution $\mathcal{N}(0,0.05)$.
The estimation result is shown in Fig.~\ref{fig:reconstruction}.
We also calculate the estimation by a standard method as used in~\cite{BelBurPil09}
with the monotone constraint only on the points $t=0,1,\dots,7$.
This figure shows that our estimation works well while the conventional one shows
oscilations in the intervals $(3,4)$ and $(5,6)$,
which indicates that the derivative is negative.
To see this more precisely, we show the derivative $\dot{y}(t)$ in Fig.~\ref{fig:derivative}.
\begin{figure}[t]
  \centering{
    \includegraphics[width=0.8\linewidth]{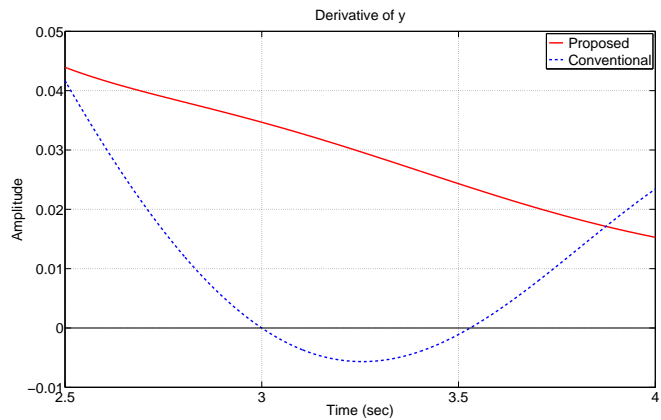}
  }
  \caption{Enlarged plot of Fig.~\ref{fig:derivative}}
  \label{fig:derivative-e}
\end{figure}
We can see that the derivative by the conventional method is non-negative 
on the sampling points $t=0,1,\dots,7$,
however the constraint is violated in the intervals $(3,4)$ and $(5,6)$
\rev{\ind{1-11} (See Fig.~\ref{fig:derivative-e} for an enlarged plot)}.
On the other hand, the derivative by our method is always non-negative 
on the whole interval $t\in[0,7]$.
\rev{\ind{3-4}
Fig.~\ref{fig:input} shows the input $u(t)$ for the both methods.
\begin{figure}[t]
 \centering
   \includegraphics[width=0.8\linewidth]{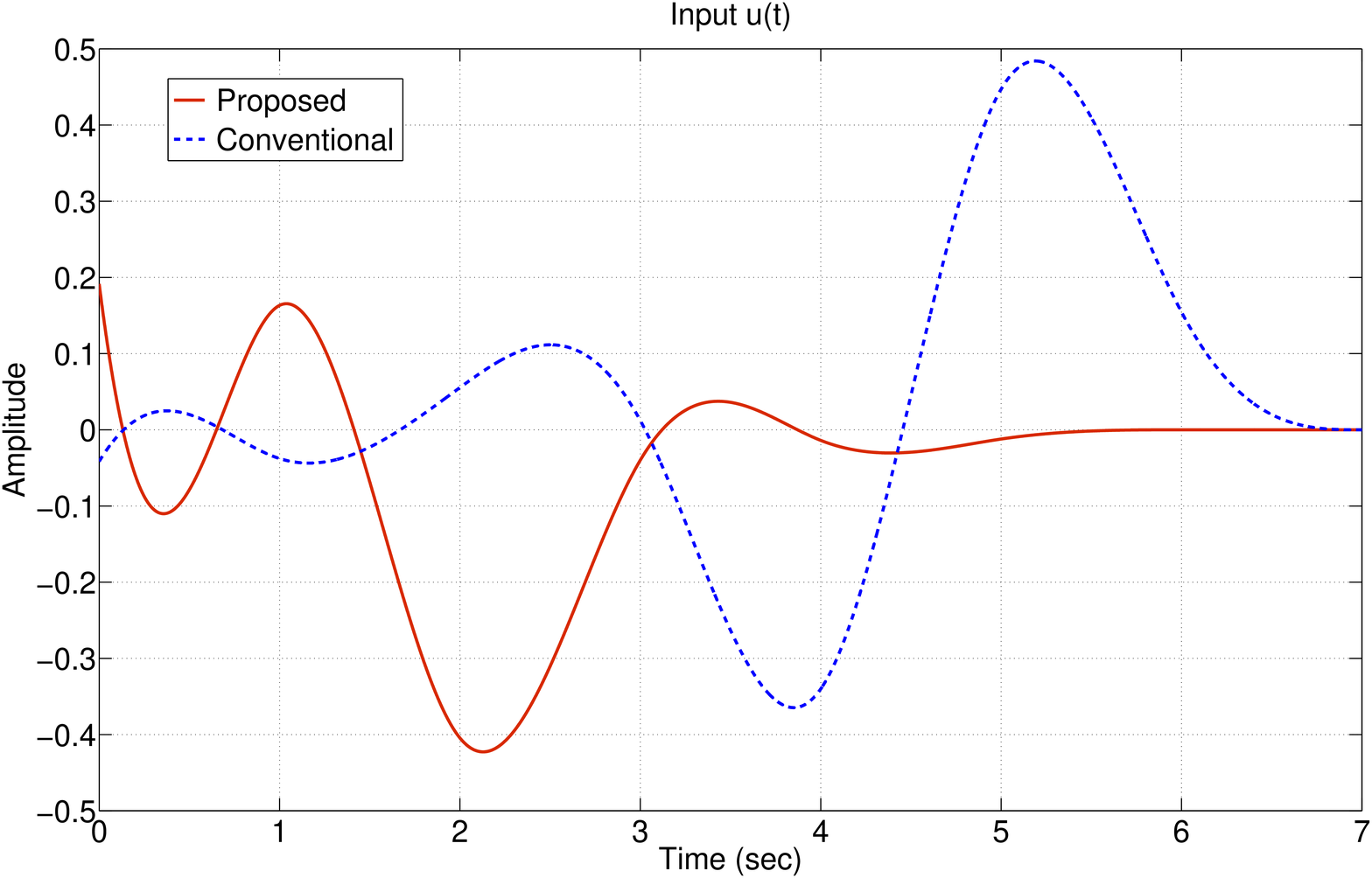}
   \caption{Input $u(t)$: proposed (solid) and conventional (dash)}
   \label{fig:input}
\end{figure}
By this figure, we can see that the input signal $u(t)$
of the conventional method shows large gain on the intervals $(3,4)$ and $(5,6)$
where the conventional method violates the monotonicity constraint.
}
\section{Conclusion}
\label{sec:conclusion}
In this paper, we have proposed a new method for solving the problem of monotone control theoretic splines
for general linear systems.
By using discretization technique, the problem is approximately described as a finite-dimensional quadratic programming.
The optimal parameters can be efficiently obtained by numerical optimization softwares such as MATLAB\@.
The obtained estimation is guaranteed to satisfy the monotonicity constraint.
We also proved limiting properties of approximated solutions.
Future work may include 
\rev{\ind{1-4}
quantification of the error introduced by the proposed approximation,
and also} construction of monotone control theoretic splines for MIMO
(multi-input multi-output) systems.

\section*{Acknowledgement}
This research was supported in part by
Grand-in-Aid for Young Scientists~(B) of 
Japan Society for the Promotion of Science under Grant No.~22760317.

\end{document}